\UseRawInputEncoding
\documentclass[aps,prl,preprint,superscriptaddress]{revtex4-2}
\usepackage{graphicx}

\begin{document}

\title{A reflective mm-wave photonic limiter}

\author{Rodion~Kononchuk}
\affiliation{Department of Physics and Astronomy, University of Texas at San Antonio, San Antonio, TX 78249, USA}
\affiliation{Department of Physics, Wesleyan University, Middletown, CT 06457, USA}

\author{Suwun~Suwunnarat}
\affiliation{Department of Physics, Wesleyan University, Middletown, CT 06457, USA}

\author{Martin~S.~Hilario}
\affiliation{Air Force Research Laboratory, Directed Energy Directorate, Kirtland AFB, NM 87117, USA}

\author{Anthony~E.~Baros}
\affiliation{Air Force Research Laboratory, Directed Energy Directorate, Kirtland AFB, NM 87117, USA}

\author{Brad~W.~Hoff}
\affiliation{Air Force Research Laboratory, Directed Energy Directorate, Kirtland AFB, NM 87117, USA}

\author{Vladimir~Vasilyev}
\affiliation{Air Force Research Laboratory, Sensors Directorate, Wright-Patterson AFB, OH 45433, USA}

\author{Ilya~Vitebskiy}
\affiliation{Air Force Research Laboratory, Sensors Directorate, Wright-Patterson AFB, OH 45433, USA}

\author{Tsampikos~Kottos}
\affiliation{Department of Physics, Wesleyan University, Middletown, CT 06457, USA}

\author{Andrey~A.~Chabanov}
\affiliation{Department of Physics and Astronomy, University of Texas at San Antonio, San Antonio, TX 78249, USA}

\date{\today}

\begin{abstract}
Millimeter wave (mm-wave) communications and radar receivers capable of processing small signals must be 
protected from high-power signals, which can damage sensitive receiver components. Many of these systems arguably 
can be protected by using photonic limiting techniques, in addition to electronic limiting circuits in receiver front-ends. 
Here we demonstrate, experimentally and numerically, a free-space, reflective mm-wave limiter based on a multilayer 
structure involving a nanolayer of vanadium dioxide (VO$_2$), experiencing a thermal insulator-to-metal transition. 
The multilayer acts as a variable reflector, controlled by the input power. At low input power levels, VO$_2$ remains 
dielectric, and the multilayer exhibits resonant transmittance. When the input power exceeds a threshold level, the 
emerging metallic phase renders the multilayer highly reflective while dissipating a small portion of the input power 
without damage to the limiter. In the case of a Gaussian beam, the limiter has a nearly constant output above 
the limiting threshold input.
\end{abstract}

\maketitle

\noindent Millimeter wave (mm-wave) is a valuable network and sensing technology which utilizes frequency bands in between 
30 GHz and 300 GHz \cite{mm-wave_comm1,mm-wave_comm2,mm-wave_sens}. Operating in this spectral range has 
important advantages over lower frequency bands. Not only do mm-waves allow larger bandwidths 
to transmit data at multi-gigabit speeds \cite{mmw_book}, they also provide higher spatial resolution due to shorter wavelengths, 
which can be exploited for a variety of accurate sensing applications \cite{TI-sensors}. Another advantage of mm-wavelengths 
is that the size of system components required to receive and process mm-wave signals is small enough to allow using standard optical techniques 
\cite{PhT,PhT1,PhT2,PhT3,PhT4,PhT5,PhT6}. 

Optical limiting is a technique to protect photosensitive devices from damage caused by intense optical radiation 
\cite{opt-lim1,opt-lim2}. Optical limiters are therefore designed to block high-intensity laser radiation while transmitting low-intensity 
light. Most passive optical limiters utilize materials with nonlinear absorption, which are transparent to low-intensity light but turn opaque 
if the light intensity exceeds a certain (limiting) threshold level \cite{opt-lim-semic,opt-lim-fuller,opt-lim-carbon}. However, a typical passive 
limiter absorbs a significant portion of high-level radiation, which can cause overheating and damage to the limiter itself \cite{opt-lim2,GaAs}. 

To overcome this problem, the concept of a reflective photonic limiter, which reflects rather than absorbs high-intensity radiation, has been 
introduced \cite{Makri1,Makri2,Vella}. A passive reflective photonic limiter involves a photonic bandgap structure, such as a multilayer cavity, 
incorporating a nonlinear \cite{Vella,SciRep} or a phase-change material (PCM) \cite{EPL,PRA}. At low incident intensities, transmittance of the 
photonic structure is at its maximum due to resonant transmission via a localized mode in photonic bandgap. High-intensity radiation, however, 
forces nonlinearity to kick in or phase transition to be induced, producing an impedance mismatch, which reflects most of the incident radiation. 
The high reflectivity prevents the limiter from overheating, thereby greatly increasing the limiter damage threshold. Other important advantages 
of the photonic design include orders-of-magnitude larger extinction ratio (the ratio of transmittances below and above the limiting threshold) 
and the possibility to significantly lower the limiting threshold by adjusting the photonic structure hosting the nonlinear material or PCM. 
The aforementioned approach thus provides advanced broadband protection from high-level radiation, although low-intensity transmission 
is essentially narrowband due to its resonant nature. Arguably, mm-wave limiters can be designed in much the same way to protect systems 
receivers from high-power signals and also to allow the receivers to function normally when these high-power signals are not present.

In this Article, we report for the first time a free-space, reflective photonic limiter for the $W$ band (75-110 GHz), inspired by recent optical designs. 
First, we design and fabricate a resonant multilayer structure incorporating a nanolayer of vanadium dioxide (VO$_2$), undergoing an insulator-to-metal 
transition when heated above the critical temperature of $\theta_c\approx69^{\circ}$C. Then, we investigate the mm-wave limiting properties of the multilayer 
by low-power spectral measurements at successively increasing temperatures and by time-resolved continuous-wave (CW) measurements at high input powers. 
Our experimental findings are then corroborated by 3D multiphysics simulations, which allow us to gain deeper insight into the limiting process by 
exploring the dynamics of the limiter driven by the VO$_2$ phase transition. 

The approach based on the use of insulator-to-metal transition materials is highly scalable and can be replicated in any spectral range, from microwave (MW) 
to optical (see, for example, \cite{MW1,MW2,Ph1,Ph2} and references therein). At MW frequencies, however, the insulator-to-metal transition is accompanied 
by much greater change in the complex permittivity of the phase-change material -- orders of magnitude in our case. This provides an unprecedented flexibility 
in control of the MW radiation flow, unattainable in optics. 

\noindent{\bf Mm-wave limiter design and low-power measurements at varying temperature.} 

\noindent The reflective mm-wave limiter studied in this work is 
illustrated in Fig.~1a,b. It involves a photonic bandgap structure constructed of 76.2-mm-diameter C-cut sapphire (Al$_2$O$_3$) wafers, with a 
525-$\mu$m thick wafer in the middle and four 256-$\mu$m thick wafers on the sides, separated by air gaps of a uniform thickness of about 
792 $\mu$m. The middle sapphire wafer is coated on one side with a polycrystalline VO$_2$ to a thickness of approximately 150 nm 
(see Methods and Supplementary Figure 1). Measurements of sheet resistance of the VO$_2$ coating reveal an abrupt insulator-to-metal 
transition at $\sim 67^{\circ}$C, with a four orders of magnitude change in the sheet resistance over a temperature range of $\sim5^{\circ}$C 
(see Supplementary Figure 2). According to literature \cite{Rozen,Basov,Hood}, in the transition region the electrical and optical properties of 
thin-film VO$_2$ are determined by the volume fraction of the thin film transformed into the metallic phase. 

\begin{figure}[ht]
\centering
\includegraphics[width= 6.0in]{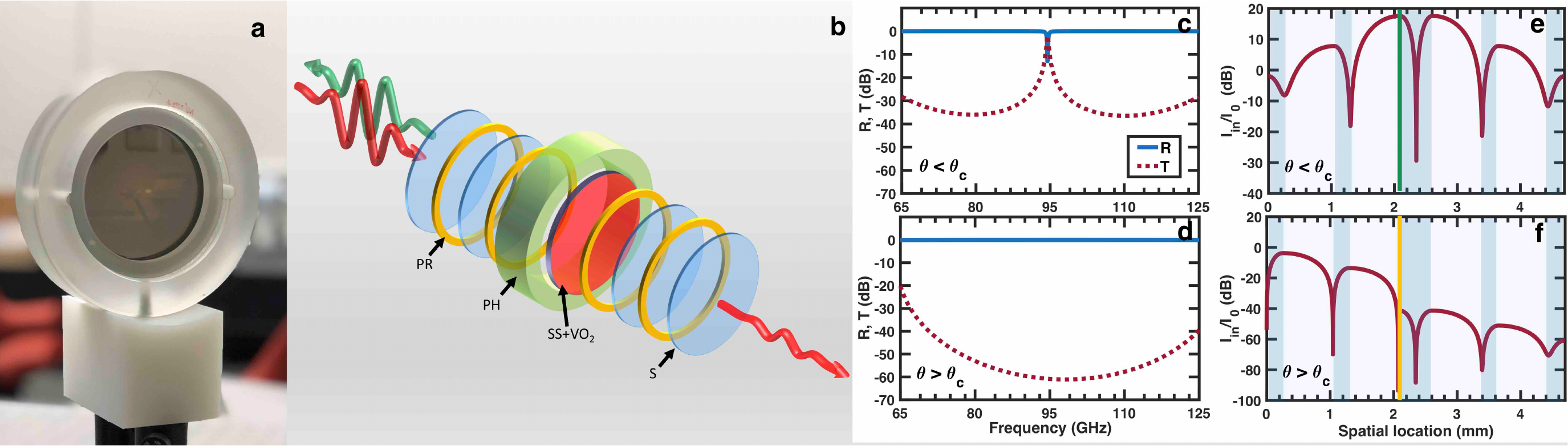}
\caption{{\bf Reflective mm-wave photonic limiter.} A picture ({\bf a}) and schematic ({\bf b}) of the mm-wave photonic limiter consisting of 
256-$\mu$m (S) and 525-$\mu$m (SS) thick sapphire wafers separated by 792-$\mu$m air-gap spacers (PR) and a 150-nm VO$_2$ layer
deposited on the SS sapphire. The layer stack is retained in a plastic holder (PH). {\bf c,d}, Simulated transmittance $T$ and reflectance $R$ 
of the photonic structure at normal incidence, at $\theta>\theta_c$ ({\bf c}) and $\theta<\theta_c$ ({\bf d}). {\bf e,f}, Simulated 
internal intensity profiles in the direction of wave propagation of the incident intensity $I_0$ at the resonance frequency 95 GHz, at 
$\theta>\theta_c$ ({\bf e}) and $\theta<\theta_c$ ({\bf f}). Sapphire and air-gap layers are shown in dark and light blue, and the 
VO$_2$ layer in green and yellow at the lower and higher temperatures, respectively.}
\label{FIG1}
\end{figure}

Since design and modeling of the photonic limiter required accurate values of the dielectric properties of the constitutive materials, we
performed mm-wave measurements of sapphire and VO$_2$-on-sapphire wafers at low power ($\leq10$ mW) and successively increased 
temperatures from 27$^{\circ}$C to 110$^{\circ}$C (see Methods). The (ordinary) relative permittivity of sapphire in the $W$ band, 
$\epsilon_{\rm s}=9.32+i0.009$, was determined from measurements of insertion loss and phase of the sapphire wafer and remained 
unchanged on heating over the whole temperature range. Fig. 2a shows the transmittance $T$ (blue circles), reflectance $R$ (black circles), 
and absorptance $A=1-T-R$ (red circles) of the VO$_2$-on-sapphire wafer at 95 GHz (in the vicinity of the Fabry-P\'erot resonance of the 
substrate) as a function of increasing temperature. The phase transition in the VO$_2$ layer manifests itself as an abrupt drop in the 
transmittance accompanied by a steep increase in the reflectance and a sharp peak in the absorptance, due to dramatic change in the 
permittivity of VO$_2$ during the transition (see Fig.~2b). The relative permittivity, $\epsilon_{\rm VO_2}=\epsilon'_{\rm VO_2}+i\epsilon''_{\rm VO_2}$, 
was found by fitting the observed insertion loss and phase of the VO$_2$-on-sapphire wafer to those obtained from numerical simulations
using $2\times 2$ transfer-matrix formalism \cite{TransferMatrix}. The real and imaginary parts, $\epsilon'_{\rm VO_2}$ and $\epsilon''_{\rm VO_2}$, 
obtained from the fit at $\theta\geq 69^{\circ}$C are plotted with the increasing temperature by solid lines (in green and magenta, respectively) 
in Fig.~2b. Below $69^{\circ}$C, the VO$_2$ contribution to the insertion loss and phase was too small to be observed; it was observed, though, 
in mm-wave measurements of the photonic limiter which follow. The corresponding $\epsilon'_{\rm VO_2}$ and $\epsilon''_{\rm VO_2}$ are 
shown by the dashed lines in Fig.~2b.

\begin{figure}[ht]
\centering
\includegraphics[width= 4.0in]{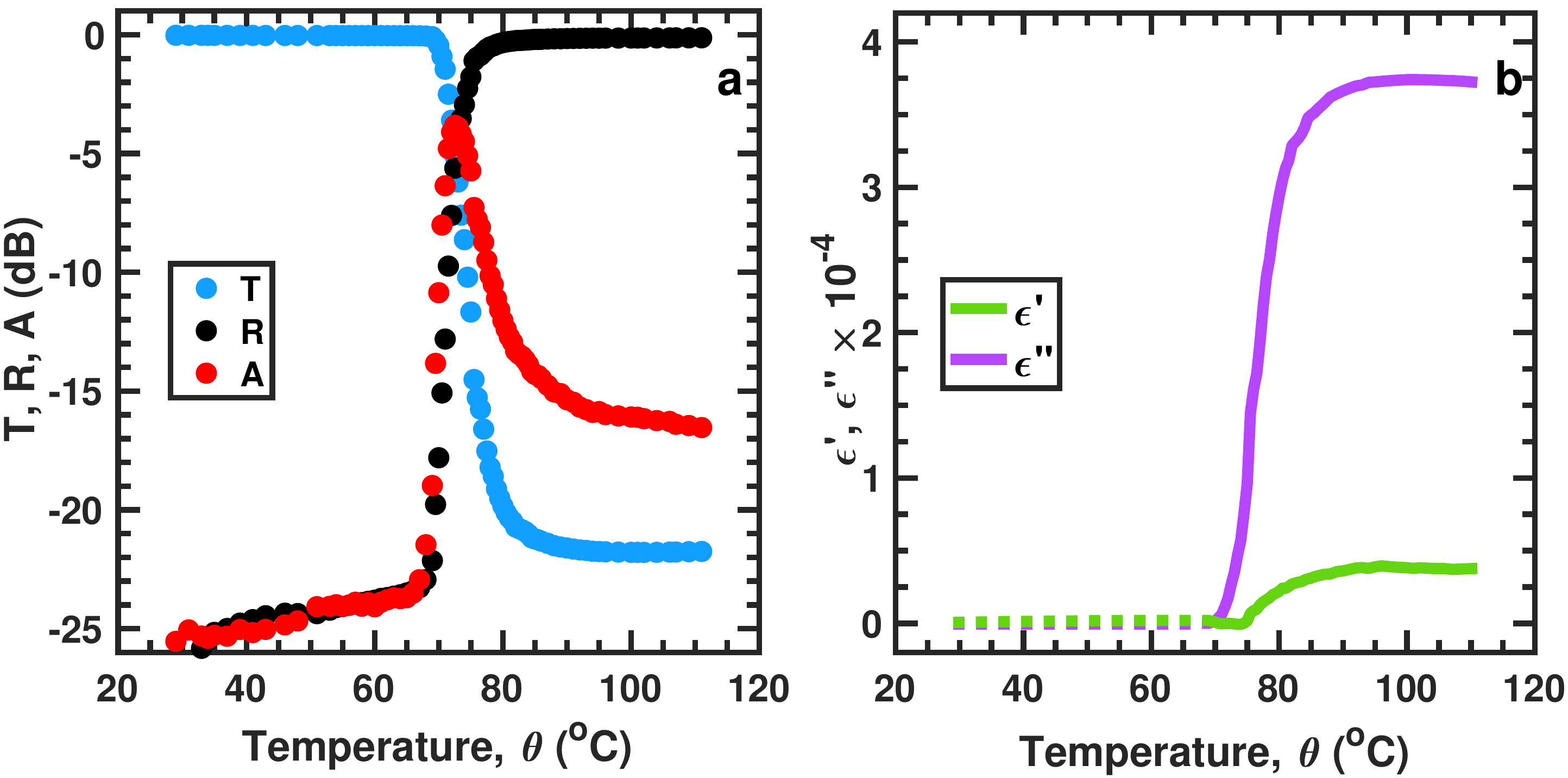}
\caption{{\bf Mm-wave dielectric properties of the VO$_2$ nanolayer at low input power and increasing temperature.} {\bf a}, Mm-wave 
transmittance $T$, reflectance $R$,  and absorptance $A$ of the VO$_2$-on-sapphire wafer at 95-GHz input power of $<10$ mW and 
successively increasing temperature $\theta$. {\bf b}, Temperature-dependent real ($\epsilon'_{\rm VO_2}$) and imaginary 
($\epsilon''_{\rm VO_2}$) parts of the relative permittivity of the VO$_2$ layer at 95 GHz, determined from the measurements of the 
insertion loss and phase of the VO$_2$-on-sapphire wafer (solid lines) and transmittance measurements of the photonic limiter (dashed lines).} 
\label{FIG2}
\end{figure}

According to Fig.~2a, the VO$_2$-on-sapphire wafer can act as a temperature-controlled mm-wave limiter with an extinction ratio of 
20 dB. As for self-induced mm-wave limiting, the threshold level of such a limiter is too high due to very low absorptance and thus weak 
heating effect at low temperatures. By using the highest (in this study) mm-wave input power (55 W) and by starting at room temperature, 
we could not heat the wafer up to $\theta_c$. 

The resonant setting of the limiter of Fig.~1a,b allows to enhance both absorptance and extinction ratio by orders of magnitude. According 
to transfer-matrix calculations below $\theta_c$, the multilayer structure exhibits a resonant transmittance (with a resonance $Q$ factor of 
260) at a frequency of the (defect-)localized mode close to 95 GHz (see Fig.~1c). Spatial intensity distribution of the electric field component 
within the multilayer at the resonance frequency is plotted with a red solid line in Fig.~1d. The position of the VO$_2$ layer (shown in green) 
is seen to coincide with an antinodal plane of the resonant electric field, providing resonance enhancement of the absorptance and thus 
heating of VO$_2$. Above $\theta_c$, the multilayer turns highly reflective over the entire $W$ band (see Fig.~1e). Moreover, the VO$_2$ 
layer (shown in orange) is shielded from incident radiation by the multilayer front-end (as seen in Fig.~1f), thus significantly increasing the 
limiter damage threshold.

Mm-wave measurements of the limiter carried out at low power and successively increased temperatures agree well with the transfer-matrix calculations. 
In Fig.~3a, the resonance transmittance peak is seen to drop by 40 dB within $70^{\circ}$C to $75^{\circ}$C temperature range, to completely 
disappear by the end of the transition. When this takes place, the limiter becomes totally reflective (see Fig.~3b). Below the transition 
temperature, the transmittance peak is seen gradually shifting to lower frequencies (see Fig.~3c), indicating an increase in the refractive 
index of VO$_2$ with the increasing temperature, while its peak value remains unchanged implying negligible VO$_2$ absorption below 
$\theta_c$. The extracted $\epsilon'_{\rm VO_2}$ and $\epsilon''_{\rm VO_2}\approx 0$ are plotted by dashed lines in Fig.~2b. Note 
that no noticeable change was observed in transmittance spectra of the multilayer without VO$_2$ coating in the same temperature range.

\begin{figure}[ht]
\centering
\includegraphics[width=6.0in]{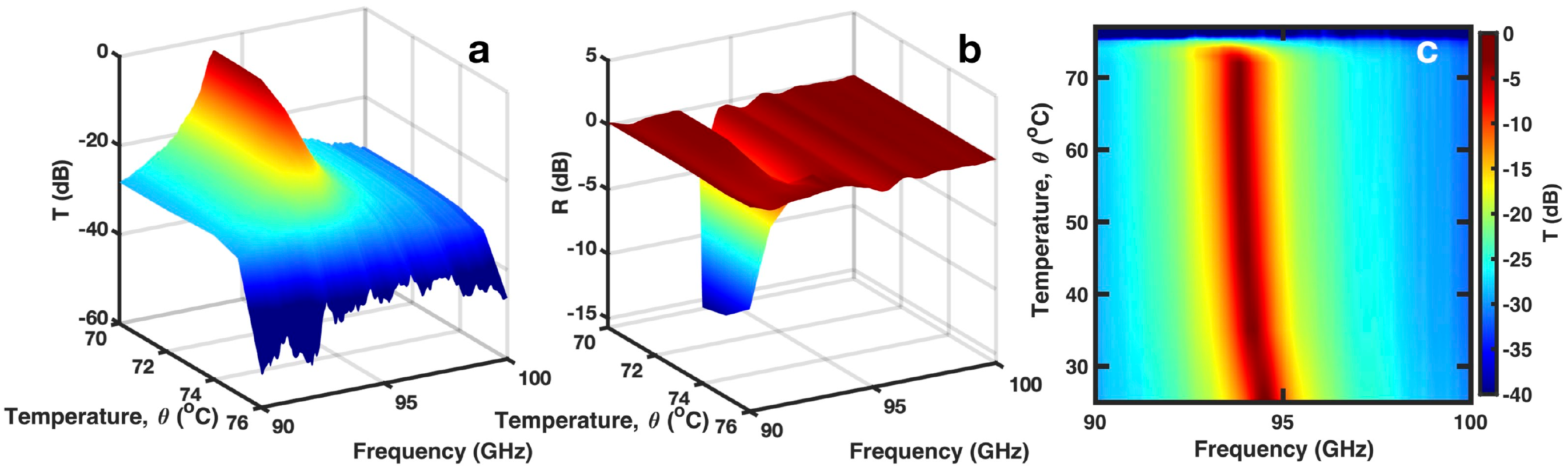}
\caption{{\bf Mm-wave measurements of the photonic limiter at low input power and increasing temperature.} {\bf a,b}, Spectra of 
transmittance $T$ ({\bf a}) and reflectance $R$ ({\bf b}) of the photonic limiter at input power of $\leq10$ mW over the resonance 
spectral range and phase-transition temperature interval. {\bf c}, Transmittance spectra over the wider temperature range from 
27$^{\circ}$C to 76$^{\circ}$C.}
\label{FIG3}
\end{figure}

\noindent{\bf High-power measurements of the mm-wave limiter.} 

\noindent To investigate self-induced mm-wave limiting, time-resolved transmission measurements of the limiter have been carried out 
with the use of a high-power CW source centered on 95 GHz with a FWHM of 300 MHz (see Methods). To prevent adverse effects of 
back-reflected radiation on the mm-wave source, the reflective limiter was tilted by an angle of $6^{\circ}$ from the normal incidence 
direction, shifting the transmission resonance to a higher frequency than 95 GHz. Fig.~4a shows linear-log plots of the transmitted 
power $P_T(t)$ for incident Gaussian beams of input powers $P_0= 30$, 35, 40, 45, and 55 W. For each input power, $P_T(t)$ is 
seen to initially increase and then decrease with time. This is due to the fact that the transmission resonance shifts to lower 
frequencies with increasing temperature, in accordance with Fig.~3c, crossing the peak frequency of 95 GHz at a time of maximum 
transmission. In addition, at input powers $P_0>30$ W, $P_T(t)$ exhibits a sharp drop associated with the insulator-to-metal transition 
in the VO$_2$ layer. The corresponding transition or switching time, $t_s$, is plotted versus $P_0$ in the inset; notice that at 30 W input 
power, the transition is not manifested in $P_T(t)$. The shortest switching time observed in these proof-of-concept measurements was 
about 7~s at the highest available input power of 55 W. It can be significantly reduced by increasing the input power (see Fig.~5a) or 
ambient temperature, and by further optimizing the limiter design.

\begin{figure}[ht]
\centering
\includegraphics[width=4.0in]{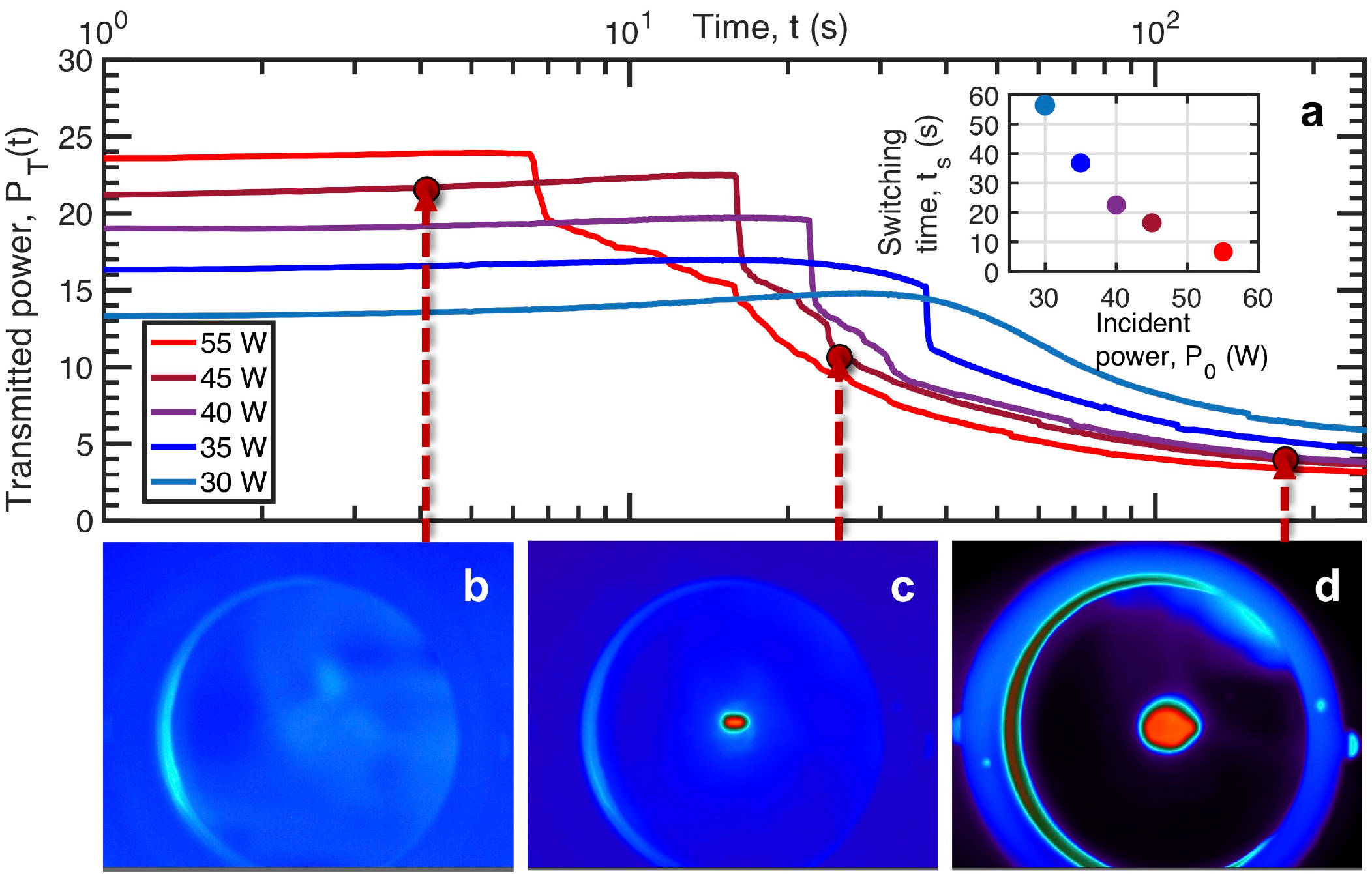}
\caption{{\bf High-power mm-wave measurements of the photonic limiter.} {\bf a}, Time-varying transmitted power $P_T(t)$ of the 
photonic limiter following excitation by a CW 95-GHz Gaussian beam of input powers $P_0=30$, 35, 40, 45, and 55 W. Inset: 
switching time $t_s$, corresponding to the onset of the metallic phase in the VO$_2$ layer, versus the input power $P_0$. {\bf b-d}, 
Thermal images of the limiter front end at elapsed times indicated by the dashed arrows in {\bf a}, for $P_0=45$ W.}
\label{FIG4}
\end{figure}

The transition in the VO$_2$ layer was observed with a FLIR thermal camera which could distinguish between the dielectric and 
metallic phases of VO$_2$ because of their different emissivities in the infrared. Fig.~4b-d display thermal images of the limiter 
front end at elapsed times indicated by dashed arrows in Fig.~4a for the incident power of 45 W. The first occurrence of the metallic 
phase in the center of the VO$_2$ layer (red spot in Fig.~4c) coincides with the sharp drop in $P_T(t)$. Then the metallic domain 
grows in size (see Fig.~4d), accompanied by further decrease in $P_T(t)$, until thermal equilibrium is reached. When high input 
power is no longer present, the VO$_2$ reverts from the metallic to the dielectric phase after a brief delay.

The equilibrium size of the metallic domain and thus the extinction ratio of the limiter increase with the incident power $P_0$. 
In contrast, the transmitted power $P_T(t)$ in the steady-state regime (i.e., at longer times) is nearly independent of  $P_0$, as seen in Fig.~4a. 
This practically important feature is, in fact, due to the Gaussian intensity distribution of the incident beam, $I(r)=(2P_0/\pi r^2_0)\exp (-2r^2/r_0^2)$. 
Indeed, if the beam is centered on and blocked by a disk of radius $a$, the transmitted power is $P_T\approx P_0\exp (-2a^2/r_0^2)$ 
provided the limiter aperture is considerably larger than the disk. Assuming that the value of $a$ is determined from $I(r=a)=I_t$, 
where $I_t$  is the intensity threshold value, the transmitted power is $P_T\approx\pi r_0^2I_t/2$ (i.e., independent of the input power $P_0$).

\noindent{\bf Electromagnetic and heat transfer modeling of the mm-wave limiter.} 

\noindent To gain deeper insight into the mm-wave limiting process in the photonic limiter, we have performed time-domain 3D 
multiphysics simulations of a Gaussian beam propagating through the multilayer of Fig.~1a,b. The electric field component 
$\mbox{\boldmath$E$}$ of the incident Gaussian beam (assuming polarization in the radial direction {\boldmath$\hat{r}$} and 
propagation in the $+z$ direction) is given by
\begin{equation}
\mbox{\boldmath$E$}(z,r) = E_0\mbox{\boldmath$\hat{r}$}\exp\!\left(-\frac{r^2}{r_0^2}\right)\exp(-ikz),
\label{e5}
\end{equation}
where $E_0$ is the on-axis ($r=0$) electric field amplitude, $E_0=\sqrt{4\eta_0 P_0/ \pi r_0^2}$, $\eta_0=377\,\Omega$ is 
the wave impedance of free space, $k = 2\pi/\lambda$ is the wave number for a free-space wavelength $\lambda$, and 
$r_0=16.5$ mm is the beam waist radius at which the field amplitude falls to $1/e$ of its axial value.

The mm-wave propagation in the case of temperature-dependent permittivity of the VO$_2$ layer is described by the 
following set of coupled electromagnetic and thermal equations:
\begin{eqnarray}
\mbox{\boldmath$\nabla$}\times\mbox{\boldmath$H$} = \mbox{\boldmath$j$}+\varepsilon\frac{\partial\mbox{\boldmath$E$}}{\partial t},\;\;\;\;
\mbox{\boldmath$\nabla$}\times\mbox{\boldmath$E$} =-\mu\frac{\partial\mbox{\boldmath$H$}}{\partial t},\nonumber\\
\rho c{\partial\theta \over \partial t}=\mbox{\boldmath$\nabla$}\cdot\left(\kappa\mbox{\boldmath$\nabla$}\theta\right)+\dot{Q}\,,\;\;\;\;\;\;\;\;\;\;\;
\label{e4}
\end{eqnarray}
where $\mbox{\boldmath$H$}$ is the magnetic field component, $\mbox{\boldmath$j$}=\sigma\mbox{\boldmath$E$}$ is the 
current density, $\sigma(z,\theta)$ is the electrical conductivity, $\varepsilon(z,\theta)=\varepsilon'(z,\theta)+i\varepsilon''(z,\theta)$ 
is the permittivity, and $\mu$ is the permeability. In the second line of equations (\ref{e4}), 
$\dot{Q} = \frac{1}{2}\left({\rm Re}(\mbox{\boldmath$j$} \cdot\mbox{\boldmath$E$}) + \omega\varepsilon''|\mbox{\boldmath$E$}|^2\right)$ 
is the volumetric heat production rate. For computational convenience, we recast the above expression as 
$\dot{Q} = \frac{1}{2} \omega\varepsilon_0\epsilon''|\mbox{\boldmath$E$}|^2$, where $\varepsilon_0$ is the vacuum permittivity, and 
$\epsilon''$ is the imaginary part of the effective relative permittivity, which can be directly obtained from our measurements. 
The other parameters in equations (\ref{e4}) include the specific heat capacity $c(z)$, the mass density $\rho(z)$, and the thermal 
conductivity $\kappa(z)$. In the modeling, we assume that the multilayer is thermally insulated on the cylindrical surface (due to the 
plastic holder) and that the heat flux $q$ across the limiter front and rear ends dissipates by convection, i.e., $q=h(\theta-\theta_{0})$, 
where $h=100$ W/(m$^2$K) is the convection heat transfer coefficient \cite{thermo}, and $\theta_{0}$ = 296 K is the ambient 
temperature. Furthermore, we use the experimentally determined relative permittivities of sapphire ($\epsilon_s$) and VO$_2$ 
($\epsilon_{\rm VO_2}$) and assume $\mu$ to be the magnetic permeability of free space, $\mu=\mu_0$. Finally, the thermal 
properties of the constitutive materials used in the simulations are listed in Supplementary Table 1.

Equations (\ref{e4}) have been solved simultaneously using the coupled Microwave and Heat Transfer modules of COMSOL 
Multiphysics software \cite{comsol}. Fig.~5 shows the computed time-dependent transmitted power $P_T(t)$, reflected power 
$P_R(t)$, and absorptance $A(t) =(P_0-P_T(t)-P_R(t))/P_0$ of the limiter following excitation by a 95-GHz Gaussian beam of 
input powers $P_{0}=29$, 56, 554, and 5540 W. A good agreement between the numerical simulations and 
the experimental data (see Fig.~4) is evident. Specifically, $P_T(t)$ exhibits abrupt, steep decrease once VO$_2$ has reached 
$\theta_c$ (see Fig.~5a). Also, in the steady limiting regime (flat portion of the curves at later times) the input power is reflected 
to a greater extent than is absorbed (see Fig.~5b,c). We further note in Fig.~5a that the switching time is inversely proportional 
to the input power, $t_{s}\propto \frac{1}{P_{0}}$, and that at $t>t_s$, the transmitted power decreases inversely with time, 
$P_T(t)\propto \frac{1}{t}$ (shown by a black dotted line).

\begin{figure}[ht]
\centering
\includegraphics[width=6in]{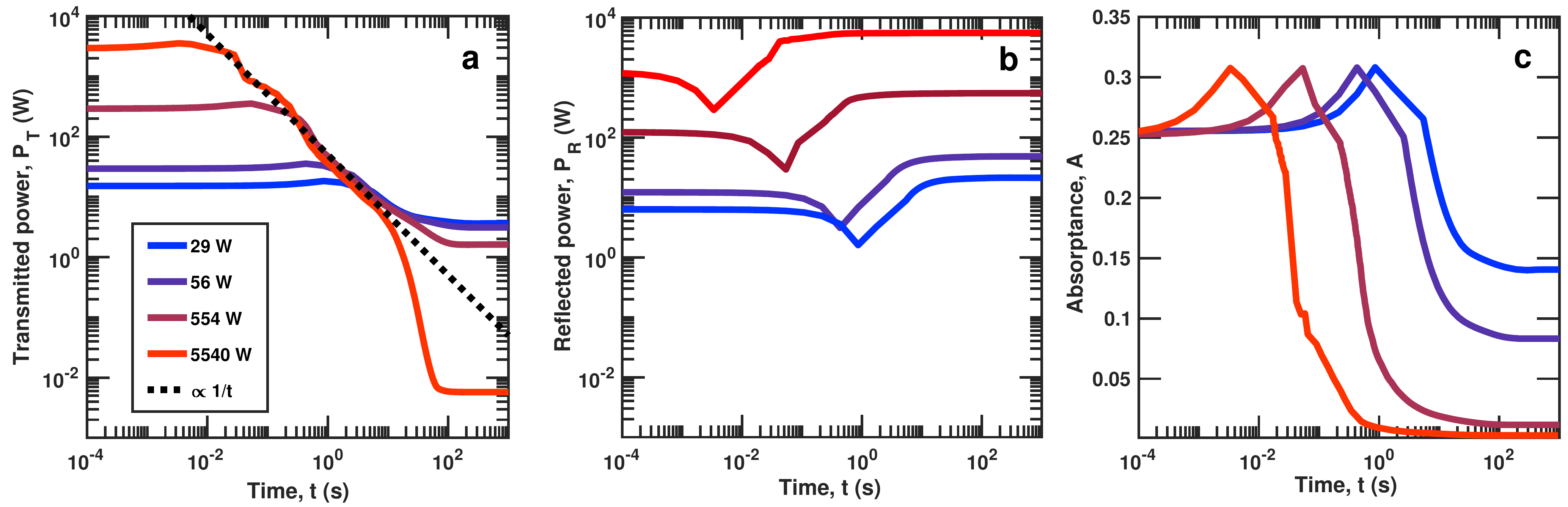}
\caption{{\bf Simulated temporal response of the photonic limiter under monochromatic excitation.} {\bf a-c}, 
Time evolution of the transmitted power $P_T(t)$ ({\bf a}), reflected power  $P_R(t)$ ({\bf b}), and absorptance $A(t)$ ({\bf c})
of the limiter following excitation by a 95-GHz Gaussian beam of input powers $P_0=29$, 56, 554, and 5540 W. 
The black dashed line in panel {\bf a} indicates a slope of $P_T(t)\propto t^{-1}$.}
\label{FIG5}
\end{figure}

The dependence of $t_{s}$ on $P_{0}$ can be understood from the solution of a heat balance equation \cite{heatbook} for the VO$_2$ layer 
in the beam center. Assuming for the time being that the temperature distribution in the VO$_2$-on-sapphire wafer is uniform 
in the $+z$ direction and neglecting transverse thermal conduction away from the beam center, the heat balance equation is
\begin{equation}
C{d\theta\over dt}=AI_{\rm in}+h'(\theta_{0}-\theta), 
\label{MEq}
\end{equation}
where $C$ is the heat capacity per unit surface area, $h'$ is the thermal exchange coefficient, $A$ is the absorptance and
 $I_{\rm in}\sim I_{0} \exp(L/2\xi)$ is the internal intensity in the middle of the multilayer length $L$, $\xi$ is the decay length 
 associated with the spatial distribution of the resonant defect mode (see Fig.~1e). Assuming that the initial temperature of 
 the VO$_2$ layer is equal to the ambient temperature, $\theta(t=0)=\theta_0$, and that $C$, $A$, and $h'$ are 
 temperature-independent below $\theta_c$, the solution of equation~(\ref{MEq}) is
\begin{equation}
\theta(t)  =\theta_{0}+{AI_{\rm in}\over h'}\left(  1-\exp\!\left(  -\frac{h'}{C}t\right)\right),
\label{CW0}
\end{equation}
for $\theta_0\leq\theta<\theta_c$. At intensities $I_{\rm in}>{h'\over A}(\theta_c-\theta_0)$, the temperature $\theta(t)$
grows until it reaches the phase transition value $\theta_c$ at time,
\begin{equation}
t_s =-{C\over h'}\ln\!\left(1-{h'\over AI_{\rm in}}(\theta_c-\theta_0)\right)\approx{C(\theta_c-\theta_0)\over AI_{\rm in}}\propto \frac{1}{P_{0}}\,.
\label{t_s}
\end{equation}

Note that in the case of the Gaussian beam of equation (1), the larger the radial distance $r$ is the longer it takes 
to reach $\theta_c$. As a result, the radius $r_m$ of the metallic domain grows with time according to 
$t={C(\theta_c-\theta_0)\over AI_{\rm in}\exp (-2r_m^2/r_0^2)}$, for $t\geq~t_s$.
Since the corresponding transmitted power is $P_T=({\pi r_0^2I_0/ 2})\exp (-2r_m^2/r_0^2)$,
it decreases with time as $P_T(t)={\pi r_0^2C(\theta_c-\theta_0)\over 2At\exp(L/2\xi)}\propto t^{-1}$, in agreement with Fig.~5a.

According to Fig.~5, it takes $\sim10^2$ s for the limiter to reach the steady state. By then $P_T(t)$ has dropped substantially 
compared to its initial level. At the input powers $P_0<60$ W, the steady-state $P_T(t)$ is independent of  $P_0$ in agreement with the 
experiment (see Fig.~4a). At $P_0>60$ W, however, $P_T(t)$ falls further down as the metallic domain becomes comparable to the 
limiter aperture in size (see Fig.~6b). At $P_0=5540$ W, $P_T(t)$ falls by 60 dB, much like in the transmittance measurement of the 
limiter at spatially uniform, successively increasing temperatures (see Fig.~3a).

As seen from Fig.~5c, the resonant setting of the multilayer enhances the initial (low-temperature) absorptance by two orders of 
magnitude compared to the stand-alone VO$_2$-on-sapphire wafer. In the presence of continuous input power $P_0$, $A(t)$ rises up
to a peak of $\sim0.32$ regardless of $P_0$ and then falls down to a constant level depending on $P_0$. This absorptance behavior, 
which is characteristic of the entire multilayer rather than the stand-alone VO$_2$-on-sapphire wafer (for comparison, see Fig.~2a), 
is essential to understanding the mm-wave limiting process in the photonic limiter. The increase in absorptance resulting from the heating effect 
triggers thermal runaway, in which the VO$_2$ temperature quickly runs up to $\theta_c$ and higher (see Fig.~6a,b), carrying VO$_2$ 
through the phase transition. However, due to rapidly increasing reflectivity with the fraction of VO$_2$ transformed 
into the metallic phase, the absorptance changes instantly from rising to falling, and the VO$_2$ layer from heating to cooling, until 
equilibrium is reached. The equilibrium sets in at a particular metallic fraction of VO$_2$, at which the limiter is capable of fully 
dissipating the absorbed power while reflecting most of the input power. If the input power further increases or decreases, so does the 
metallic fraction of VO$_2$ to reset the equilibrium. In contrast, the stand-alone VO$_2$-on-sapphire wafer exhibits only thermal runaway 
and lacks equilibrium at  $\theta>69^{\circ}$C, because its absorbance at $\theta\geq\theta_c$ is higher than that at $\theta<\theta_c$ 
(see Fig.~2a).

\begin{figure}[ht]
\centering
\includegraphics[width= 4in]{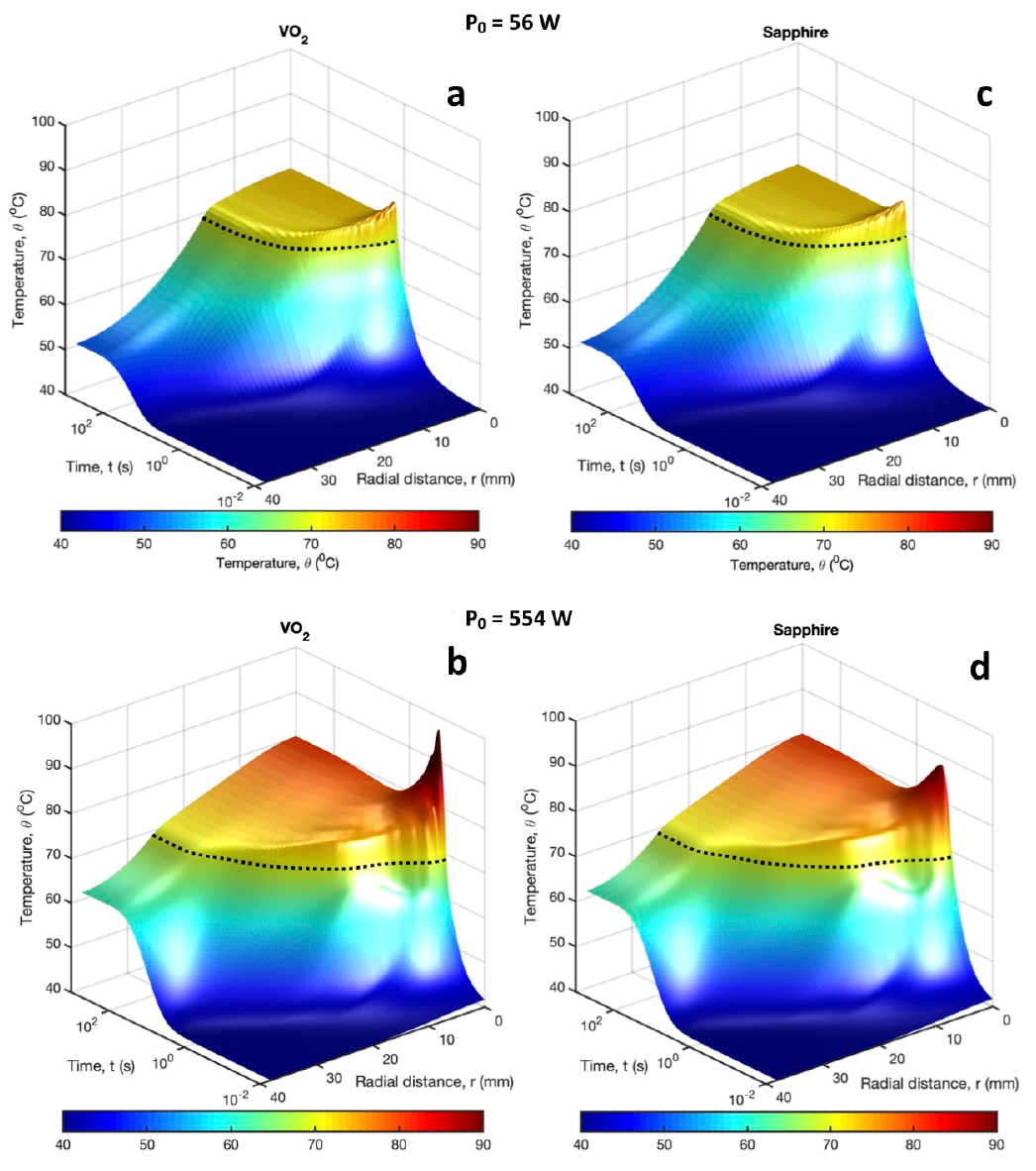}
\caption{{\bf Simulated temperature evolution of the VO$_2$-on-sapphire wafer inside the photonic limiter under monochromatic excitation.} 
{\bf a,b}, Radial temperature distribution of the VO$_2$ layer as a function of elapsed time $t$ following excitation by 
a 95-GHz Gaussian beam with the waist radius $r_0=16.5$ mm and input powers of 56 W  ({\bf a}) and 554 W  ({\bf b}). 
{\bf c,d}, the same for the opposite (uncoated) side of the sapphire wafer. The black dashed lines correspond to 
$\theta=69^{\circ}$C.}
\label{FIG6}
\end{figure}

The picture of a self-regulating limiting process in the photonic limiter is further supported by the computed temperature time evolution 
of the VO$_2$-on-sapphire wafer inside the photonic limiter at input powers of 56 W and 554 W shown in Fig.~6. Due to the limiter's axial symmetry, we have 
plotted time-varying radial temperature distributions at the two $z$-positions: in the VO$_2$ layer (see Fig.~6a,b) and on the opposite 
side of the sapphire wafer (see Fig.~6c,d). Below $69^{\circ}$C (shown by a black dashed line), the temperature distributions at these 
two positions are nearly the same, indicating that the VO$_2$ layer (assumed to be lossless) is efficiently heated by the absorbing 
sapphire wafer. At $\theta>69^{\circ}$C, the VO$_2$ temperature rises faster and higher than in sapphire, particularly at the beam 
center ($r=0$) and the higher power of 554 W. The two temperatures peak and fall at the same time despite a temperature gradient 
across the sapphire wafer, indicating that sapphire is shielded by VO$_2$ from the incident beam as the latter is being transformed 
into the metallic phase. Next, the two temperatures quickly level off, eventually reaching equilibrium temperatures of $75^{\circ}$C 
and $80^{\circ}$C for input powers of 56W and 554W, respectively, in accordance with Fig.~5 and the subsequent discussion.

\noindent {\bf Conclusion}

\noindent We have designed and demonstrated a free-space, reflective photonic limiter for the $W$ band as an alternative or a complement 
to electronic limiting circuits used in mm-wave receivers. The proposed photonic limiter is a resonant air/sapphire multilayer structure, 
a few mm thick, incorporating a 150-nm VO$_2$ layer undergoing an insulator-to-metal transition when heated above the critical 
temperature, $\theta_c\approx69^{\circ}$C. The limiter acts as a variable reflector, controlled by the input power $P_0$. If $P_0$ is 
below a certain threshold level $P_t$, the VO$_2$ layer remains in the dielectric state, and the multilayer exhibits high transmittance 
in a finite frequency band, despite the fact that a portion of $P_0$ is absorbed by the multilayer, leading to some heating of VO$_2$. 
When $P_0$ exceeds $P_t$, a fraction of the VO$_2$ layer transitions into the metallic phase with sharply increased electrical 
conductivity, rendering the entire multilayer highly reflective. The time $t_s$ before the transition starts is inversely proportional to 
the input power, $t_{s}\propto \frac{1}{P_{0}}$, after which the multilayer is brought to equilibrium by a combination of 
positive and negative feedbacks between mm-wave absorption and heating of VO$_2$. For a given input power $P_0\geq P_t$, 
the equilibrium is determined by the fraction of the VO$_2$ layer transformed into the metallic phase. At that point, the limiter is 
capable of safely dissipating a small portion of the incident power while reflecting the rest of the energy back to space. The high 
reflectivity (as opposed to absorptance) prevents the limiter from overheating. Moreover, a combination of the resonant conditions 
and the high contrast in electrical conductivity between the two phases of VO$_2$ results in a significant enhancement of the 
limiter extinction ratio and allows drastic reduction in the limiting threshold --- both highly desirable features in limiting 
applications. 

The limiter properties, such as operating frequency, bandwidth, threshold level and switching time, are determined by 
the limiter design, ambient conditions, as well as spatial and temporal characteristics of the incident radiation. In particular, 
in the case of CW Gaussian beam, the output power of the limiter operating above the threshold appears finite and virtually 
independent of the input power due to non-uniform heating of the VO$_2$ layer.

\noindent {\bf Acknowledgement}

\noindent This research has been supported by the Air Force Office of Scientific Research (FA9550-19-1-0359, LRIR 18RYCOR013, 
LRIR 20RDCOR022), Office of Naval Research (N00014-19-1-2480), and DARPA (HR00111820042). We thank K. Leedy and E. Shin 
for technical help with the PLD system and helpful discussions.

\noindent {\bf Methods}

\noindent{\bf Thin-film deposition and structural analysis.} The VO$_2$ thin film was deposited in a 
Neocera Pioneer 180 Pulse Laser Deposition system with a KrF excimer laser (Coherent COMPexPro 110, 
$\lambda = 248$ nm, 10-ns pulse duration, 10-Hz repetition rate) applied for the ablation of a high-purity 
vanadium disk. The chamber base pressure of a 5\%-O2/95\%-Ar gas mixture was maintained at 25 mTorr 
during the deposition. The 76.2-mm double-side-polished C-cut (006) sapphire substrate was held at 600$^{\circ}$C.
X-ray diffraction (XRD) analysis of the VO$_2$ films was carried out using a PANalytical X-Pert diffractometer 
with a hybrid monochromator for Cu K$\alpha_1$ radiation ($\lambda = 1.54056$ \AA) at room temperature. 
The presence of highly oriented monoclinic VO$_2$ crystalline phase is indicated by the characteristic 
XRD peaks (020) and (040) at $39.87^{\circ}$ and $85.97^{\circ}$, respectively \cite{XRD1} (see Supplementary 
Figure 1). The average VO$_2$ crystalline grain size about 110 nm was found from the Scherrer equation 
for the FWHM ($\beta$) of the (020) peak \cite{XRD2}. A VO$_2$ film thickness of approximately 150 nm was 
measured with SEM.

\noindent{\bf Mm-wave spectral measurements.} Mm-wave field spectra of the photonic limiter and complex 
permittivities of its constitutive components were obtained using a $W$-band, high-temperature, free-space 
measurement apparatus \cite{RD}. The measurement system consisted of an Agilent 5222A performance 
network analyzer (PNA), N5261A millimeter head controller, and a set of OML V10VNA-T/R frequency extender 
heads (one for each port) serving to boost the output of the PNA base unit to $W$-band frequencies (75--110 GHz). 
A matched set of custom-designed, lensed horn antennas was used for launching and receiving a Gaussian mm-wave 
beam (waist radius of 16.5 mm) transmitted via and reflected from the sample. The sample was located in the center 
of a Mellen tube furnace with an overall length and diameter of 31.75 and 15.24 cm, respectively, and a 15.24-cm-long, 
uniform heating region centered along the axial length of the furnace. A silicon carbide composite sample holder was 
used for the measurements. The temperature was controlled with a E-type thermocouple embedded in the sample 
holder immediately adjacent to the outer radius of the sample. In the temperature range from $27^{\circ}$C to 
$110^{\circ}$C, the thermocouple had an accuracy of $1.0^{\circ}$C.

\noindent{\bf High-power mm-wave measurements.} For higher-power free-space measurements, a CPI 
VKB2463L2 Extended Interaction Klystron (EIK) power amplifier was used to provide a CW 95-GHz signal. 
EIK input and output mm-wave power was sampled utilizing calibrated directional couplers, and then measured 
using diode-based detectors. The input mm-wave drive signal was controlled with a LabView-based feedback 
circuit to maintain a consistent mm-wave power output over extended periods of time, which could otherwise 
drop due to heating of components. The EIK mm-wave output was coupled into the same lensed antenna 
that was utilized with the network analyzer measurement setup, to form a Gaussian beam directed into a 
shielded anechoic chamber and focused at the center of the sample being interrogated. At a peak power 
gain, the maximum incident power of the Gaussian mm-wave beam was 55 W.

\noindent{\bf Multiphysics simulations.} Equations (2) were solved numerically, using the coupled Microwave 
and Heat transfer modules of a finite-element software package from COMSOL MULTIPHYSICS \cite{comsol} 
and a 2D axisymmetric model, capturing all the features of the 3D problem with axial (cylindrical) symmetry. 
In the simulations, we used a varying mesh density (see Supplementary Figure 3). The thin sapphire and air-gap 
layers were partitioned with the mesh density increased towards the center of the limiter, while the center 
sapphire and air-gap layers towards the VO$_2$ layer. The convergence of the results has been evaluated with 
a tolerance factor of 0.1\%. We then repeated the calculations by doubling the number of mesh points in order 
to guarantee the accuracy of the converged numerical solutions. The mm-wave source was modeled with a surface 
current density line producing the Gaussian beam, which was located 1.5 mm away from the front end of the multilayer 
inside the far-field air domain. The wave propagation was modeled using PML boundary conditions to avoid the back 
reflections from the boundaries of the geometry.

\end{document}